
%
%
%
%
%
%
%
%
%
\def\standardrisposta{s }\def\reducedrisposta{r }
\def\mplarisposta{mpla }\def\zerorisposta{z }
\def\doublerisposta{d }\def\cartarisposta{e }\def\amsrisposta{y }
\newcount\ingrandimento \newcount\sinnota \newcount\dimnota
\newcount\unoduecol \newdimen\collhsize \newdimen\tothsize
\newdimen\fullhsize \newcount\controllorisposta \sinnota=1
\newskip\infralinea  \global\controllorisposta=0
\immediate\write16 { ********  Welcome to PANDA macros (Plain TeX,
AP, 1991) ******** }
\immediate\write16 { You'll have to answer a few questions in
lowercase.}
\message{>  Do you want it in double-page (d), reduced (r)
or standard format (s) ? }\read-1 to\risposta
\message{>  Do you want it in USA A4 (u) or EUROPEAN A4
(e) paper size ? }\read-1 to\srisposta
\message{>  Do you have AMSFonts 2.0 (math) fonts (y/n) ? }
\read-1 to\arisposta
%
%
%
%
%
\ifx\risposta\standardrisposta \ingrandimento=1200
\message {>> This will come out UNREDUCED << }
\dimnota=2 \unoduecol=1 \global\controllorisposta=1 \fi
\ifx\risposta\reducedrisposta \ingrandimento=1095 \dimnota=1
\unoduecol=1  \global\controllorisposta=1
\message {>> This will come out REDUCED << } \fi
\ifx\risposta\doublerisposta \ingrandimento=1000 \dimnota=2
\unoduecol=2  \message {>> You must print this in
LANDSCAPE orientation << } \global\controllorisposta=1 \fi
\ifx\risposta\mplarisposta \ingrandimento=1000 \dimnota=1
\message {>> Mod. Phys. Lett. A format << }
\unoduecol=1 \global\controllorisposta=1 \fi
\ifx\risposta\zerorisposta \ingrandimento=1000 \dimnota=2
\message {>> Zero Magnification format << }
\unoduecol=1 \global\controllorisposta=1 \fi
\ifnum\controllorisposta=0  \ingrandimento=1200
\message {>>> ERROR IN INPUT, I ASSUME STANDARD
UNREDUCED FORMAT <<< }  \dimnota=2 \unoduecol=1 \fi
\magnification=\ingrandimento
%
%
%
%
\newdimen\eucolumnsize \newdimen\eudoublehsize \newdimen\eudoublevsize
\newdimen\uscolumnsize \newdimen\usdoublehsize \newdimen\usdoublevsize
\newdimen\eusinglehsize \newdimen\eusinglevsize \newdimen\ussinglehsize
\newskip\standardbaselineskip \newdimen\ussinglevsize
\newskip\reducedbaselineskip \newskip\doublebaselineskip
\eucolumnsize=12.0truecm    
\eudoublehsize=25.5truecm   
\eudoublevsize=6.5truein    
\uscolumnsize=4.4truein     
\usdoublehsize=9.4truein    
\usdoublevsize=6.8truein    
\eusinglehsize=6.3truein    
\eusinglevsize=24truecm     
\ussinglehsize=6.5truein    
\ussinglevsize=8.9truein    
\standardbaselineskip=16pt plus.2pt  
\reducedbaselineskip=14pt plus.2pt   
\doublebaselineskip=12pt plus.2pt    
%
%
\def\Portoffset{}
\def\Landoffset{\hoffset=-.140truein}
\ifx\risposta\mplarisposta \def\Portoffset{\hoffset=1.9truecm
\voffset=1.4truecm} \fi
%
%
\def\Landspec{}
\tolerance=10000
\parskip=0pt plus2pt  \leftskip=0pt \rightskip=0pt
%
%
\ifx\risposta\standardrisposta \infralinea=\standardbaselineskip \fi
\ifx\risposta\reducedrisposta  \infralinea=\reducedbaselineskip \fi
\ifx\risposta\doublerisposta   \infralinea=\doublebaselineskip \fi
\ifx\risposta\mplarisposta     \infralinea=13pt \fi
\ifx\risposta\zerorisposta     \infralinea=12pt plus.2pt\fi
\ifnum\controllorisposta=0    \infralinea=\standardbaselineskip \fi
\ifx\risposta\doublerisposta   \Landoffset \else \Portoffset \fi
\ifx\risposta\doublerisposta \ifx\srisposta\cartarisposta
\tothsize=\eudoublehsize \collhsize=\eucolumnsize
\vsize=\eudoublevsize  \else  \tothsize=\usdoublehsize
\collhsize=\uscolumnsize \vsize=\usdoublevsize \fi \else
\ifx\srisposta\cartarisposta \tothsize=\eusinglehsize
\vsize=\eusinglevsize \else  \tothsize=\ussinglehsize
\vsize=\ussinglevsize \fi \collhsize=4.4truein \fi
\ifx\risposta\mplarisposta \tothsize=5.0truein
\vsize=7.8truein \collhsize=4.4truein \fi
%
%
%
%
\newcount\contaeuler \newcount\contacyrill \newcount\contaams
\font\ninerm=cmr9  \font\eightrm=cmr8  \font\sixrm=cmr6
\font\ninei=cmmi9  \font\eighti=cmmi8  \font\sixi=cmmi6
\font\ninesy=cmsy9  \font\eightsy=cmsy8  \font\sixsy=cmsy6
\font\ninebf=cmbx9  \font\eightbf=cmbx8  \font\sixbf=cmbx6
\font\ninett=cmtt9  \font\eighttt=cmtt8  \font\nineit=cmti9
\font\eightit=cmti8 \font\ninesl=cmsl9  \font\eightsl=cmsl8
\skewchar\ninei='177 \skewchar\eighti='177 \skewchar\sixi='177
\skewchar\ninesy='60 \skewchar\eightsy='60 \skewchar\sixsy='60
\hyphenchar\ninett=-1 \hyphenchar\eighttt=-1 \hyphenchar\tentt=-1
%
\font\tencmmib=cmmib10  \newfam\cmmibfam  \skewchar\tencmmib='177
\font\tencmbsy=cmbsy10  \newfam\cmbsyfam  \skewchar\tencmbsy='60
\def\scaps{\cmcsc}                 
\font\tencmcsc=cmcsc10  \newfam\cmcscfam
\ifnum\ingrandimento=1095 
 
\font\bfone=cmbx10 at 10.95pt

\font\capsone=cmcsc10 at 10.95pt 

\else  
 
\font\bfone=cmbx10 at 12pt

\font\capsone=cmcsc10 at 12pt 
\fi
\def\chapterfont#1{\xdef\ttaarr{#1}}
\def\sectionfont#1{\xdef\ppaarr{#1}}
\def\ttaarr{\bf}		
\def\ppaarr{\sl}		

%
%
%
\newfam\eufmfam \newfam\msamfam \newfam\msbmfam \newfam\eufbfam
\def\loadamsmath{\global\contaams=1 \ifx\arisposta\amsrisposta
\font\tenmsam=msam10 \font\ninemsam=msam9 \font\eightmsam=msam8
\font\sevenmsam=msam7 \font\sixmsam=msam6 \font\fivemsam=msam5
\font\tenmsbm=msbm10 \font\ninemsbm=msbm9 \font\eightmsbm=msbm8
\font\sevenmsbm=msbm7 \font\sixmsbm=msbm6 \font\fivemsbm=msbm5
\else \def\msbm{\bf} \fi \def\Bbb{\msbm} \def\symbl{\msam} \tenpoint}
\ifx\arisposta\amsrisposta
\font\sevenex=cmex7               
\font\eightex=cmex8  \font\nineex=cmex9
\font\ninecmmib=cmmib9   \font\eightcmmib=cmmib8
\font\sevencmmib=cmmib7 \font\sixcmmib=cmmib6
\font\fivecmmib=cmmib5   \skewchar\ninecmmib='177
\skewchar\eightcmmib='177  \skewchar\sevencmmib='177
\skewchar\sixcmmib='177   \skewchar\fivecmmib='177
\font\ninecmbsy=cmbsy9    \font\eightcmbsy=cmbsy8
\font\sevencmbsy=cmbsy7  \font\sixcmbsy=cmbsy6
\font\fivecmbsy=cmbsy5   \skewchar\ninecmbsy='60
\skewchar\eightcmbsy='60  \skewchar\sevencmbsy='60
\skewchar\sixcmbsy='60    \skewchar\fivecmbsy='60
\font\ninecmcsc=cmcsc9    \font\eightcmcsc=cmcsc8     \else
\def\cmmib{\fam\cmmibfam\tencmmib}\textfont\cmmibfam=\tencmmib
\scriptfont\cmmibfam=\tencmmib \scriptscriptfont\cmmibfam=\tencmmib
\def\cmbsy{\fam\cmbsyfam\tencmbsy} \textfont\cmbsyfam=\tencmbsy
\scriptfont\cmbsyfam=\tencmbsy \scriptscriptfont\cmbsyfam=\tencmbsy
\scriptfont\cmcscfam=\tencmcsc \scriptscriptfont\cmcscfam=\tencmcsc
\def\cmcsc{\fam\cmcscfam\tencmcsc} \textfont\cmcscfam=\tencmcsc \fi
\catcode`@=11
\newskip\ttglue
\gdef\tenpoint{\def\rm{\fam0\tenrm}
  \textfont0=\tenrm \scriptfont0=\sevenrm \scriptscriptfont0=\fiverm
  \textfont1=\teni \scriptfont1=\seveni \scriptscriptfont1=\fivei
  \textfont2=\tensy \scriptfont2=\sevensy \scriptscriptfont2=\fivesy
  \textfont3=\tenex \scriptfont3=\tenex \scriptscriptfont3=\tenex
  \def\mcal{\fam2 \tensy}  \def\mmit{\fam1 \teni}
  \textfont\itfam=\tenit \def\it{\fam\itfam\tenit}
  \textfont\slfam=\tensl \def\sl{\fam\slfam\tensl}
  \textfont\ttfam=\tentt \scriptfont\ttfam=\eighttt
  \scriptscriptfont\ttfam=\eighttt  \def\tt{\fam\ttfam\tentt}
  \textfont\bffam=\tenbf \scriptfont\bffam=\sevenbf
  \scriptscriptfont\bffam=\fivebf \def\bf{\fam\bffam\tenbf}
     \ifx\arisposta\amsrisposta      \ifnum\contaams=1
  \textfont\msamfam=\tenmsam \scriptfont\msamfam=\sevenmsam
  \scriptscriptfont\msamfam=\fivemsam \def\msam{\fam\msamfam\tenmsam}
  \textfont\msbmfam=\tenmsbm \scriptfont\msbmfam=\sevenmsbm
  \scriptscriptfont\msbmfam=\fivemsbm \def\msbm{\fam\msbmfam\tenmsbm}\fi
  \textfont3=\tenex \scriptfont3=\sevenex \scriptscriptfont3=\sevenex
  \def\cmmib{\fam\cmmibfam\tencmmib} \scriptfont\cmmibfam=\sevencmmib
  \textfont\cmmibfam=\tencmmib  \scriptscriptfont\cmmibfam=\fivecmmib
  \def\cmbsy{\fam\cmbsyfam\tencmbsy} \scriptfont\cmbsyfam=\sevencmbsy
  \textfont\cmbsyfam=\tencmbsy  \scriptscriptfont\cmbsyfam=\fivecmbsy
  \def\cmcsc{\fam\cmcscfam\tencmcsc} \scriptfont\cmcscfam=\eightcmcsc
  \textfont\cmcscfam=\tencmcsc \scriptscriptfont\cmcscfam=\eightcmcsc
     \fi            \tt \ttglue=.5em plus.25em minus.15em
  \normalbaselineskip=12pt
  \setbox\strutbox=\hbox{\vrule height8.5pt depth3.5pt width0pt}
  \let\sc=\eightrm \let\big=\tenbig   \normalbaselines
  \baselineskip=\infralinea  \rm}
\gdef\ninepoint{\def\rm{\fam0\ninerm}
  \textfont0=\ninerm \scriptfont0=\sixrm \scriptscriptfont0=\fiverm
  \textfont1=\ninei \scriptfont1=\sixi \scriptscriptfont1=\fivei
  \textfont2=\ninesy \scriptfont2=\sixsy \scriptscriptfont2=\fivesy
  \textfont3=\tenex \scriptfont3=\tenex \scriptscriptfont3=\tenex
  \def\mcal{\fam2 \ninesy}  \def\mmit{\fam1 \ninei}
  \textfont\itfam=\nineit \def\it{\fam\itfam\nineit}
  \textfont\slfam=\ninesl \def\sl{\fam\slfam\ninesl}
  \textfont\ttfam=\ninett \scriptfont\ttfam=\eighttt
  \scriptscriptfont\ttfam=\eighttt \def\tt{\fam\ttfam\ninett}
  \textfont\bffam=\ninebf \scriptfont\bffam=\sixbf
  \scriptscriptfont\bffam=\fivebf \def\bf{\fam\bffam\ninebf}
     \ifx\arisposta\amsrisposta   \ifnum\contaams=1
  \textfont\msamfam=\ninemsam \scriptfont\msamfam=\sixmsam
  \scriptscriptfont\msamfam=\fivemsam \def\msam{\fam\msamfam\ninemsam}
  \textfont\msbmfam=\ninemsbm \scriptfont\msbmfam=\sixmsbm
  \scriptscriptfont\msbmfam=\fivemsbm \def\msbm{\fam\msbmfam\ninemsbm}\fi
  \textfont3=\nineex \scriptfont3=\sevenex \scriptscriptfont3=\sevenex
  \def\cmmib{\fam\cmmibfam\ninecmmib}  \textfont\cmmibfam=\ninecmmib
  \scriptfont\cmmibfam=\sixcmmib \scriptscriptfont\cmmibfam=\fivecmmib
  \def\cmbsy{\fam\cmbsyfam\ninecmbsy}  \textfont\cmbsyfam=\ninecmbsy
  \scriptfont\cmbsyfam=\sixcmbsy \scriptscriptfont\cmbsyfam=\fivecmbsy
  \def\cmcsc{\fam\cmcscfam\ninecmcsc} \scriptfont\cmcscfam=\eightcmcsc
  \textfont\cmcscfam=\ninecmcsc \scriptscriptfont\cmcscfam=\eightcmcsc
     \fi            \tt \ttglue=.5em plus.25em minus.15em
  \normalbaselineskip=11pt
  \setbox\strutbox=\hbox{\vrule height8pt depth3pt width0pt}
  \let\sc=\sevenrm \let\big=\ninebig \normalbaselines\rm}
\gdef\eightpoint{\def\rm{\fam0\eightrm}
  \textfont0=\eightrm \scriptfont0=\sixrm \scriptscriptfont0=\fiverm
  \textfont1=\eighti \scriptfont1=\sixi \scriptscriptfont1=\fivei
  \textfont2=\eightsy \scriptfont2=\sixsy \scriptscriptfont2=\fivesy
  \textfont3=\tenex \scriptfont3=\tenex \scriptscriptfont3=\tenex
  \def\mcal{\fam2 \eightsy}  \def\mmit{\fam1 \eighti}
  \textfont\itfam=\eightit \def\it{\fam\itfam\eightit}
  \textfont\slfam=\eightsl \def\sl{\fam\slfam\eightsl}
  \textfont\ttfam=\eighttt \scriptfont\ttfam=\eighttt
  \scriptscriptfont\ttfam=\eighttt \def\tt{\fam\ttfam\eighttt}
  \textfont\bffam=\eightbf \scriptfont\bffam=\sixbf
  \scriptscriptfont\bffam=\fivebf \def\bf{\fam\bffam\eightbf}
     \ifx\arisposta\amsrisposta      \ifnum\contaams=1
  \textfont\msamfam=\eightmsam \scriptfont\msamfam=\sixmsam
  \scriptscriptfont\msamfam=\fivemsam \def\msam{\fam\msamfam\eightmsam}
  \textfont\msbmfam=\eightmsbm \scriptfont\msbmfam=\sixmsbm
  \scriptscriptfont\msbmfam=\fivemsbm \def\msbm{\fam\msbmfam\eightmsbm}\fi
  \textfont3=\eightex \scriptfont3=\sevenex \scriptscriptfont3=\sevenex
  \def\cmmib{\fam\cmmibfam\eightcmmib}  \textfont\cmmibfam=\eightcmmib
  \scriptfont\cmmibfam=\sixcmmib \scriptscriptfont\cmmibfam=\fivecmmib
  \def\cmbsy{\fam\cmbsyfam\eightcmbsy}  \textfont\cmbsyfam=\eightcmbsy
  \scriptfont\cmbsyfam=\sixcmbsy \scriptscriptfont\cmbsyfam=\fivecmbsy
  \def\cmcsc{\fam\cmcscfam\eightcmcsc} \scriptfont\cmcscfam=\eightcmcsc
  \textfont\cmcscfam=\eightcmcsc \scriptscriptfont\cmcscfam=\eightcmcsc
     \fi             \tt \ttglue=.5em plus.25em minus.15em
  \normalbaselineskip=9pt
  \setbox\strutbox=\hbox{\vrule height7pt depth2pt width0pt}
  \let\sc=\sixrm \let\big=\eightbig \normalbaselines\rm }
\global\contaeuler=0 \global\contacyrill=0 \global\contaams=0
%
%
%
%
\newbox\fotlinebb \newbox\hedlinebb \newbox\leftcolumn
\gdef\makeheadline{\vbox to 0pt{\vskip-22.5pt
     \fullline{\vbox to8.5pt{}\the\headline}\vss}\nointerlineskip}
\gdef\makehedlinebb{\vbox to 0pt{\vskip-22.5pt
     \fullline{\vbox to8.5pt{}\copy\hedlinebb\hfil
     \line{\hfill\the\headline\hfill}}\vss} \nointerlineskip}
\gdef\makefootline{\baselineskip=24pt \fullline{\the\footline}}
\gdef\makefotlinebb{\baselineskip=24pt
    \fullline{\copy\fotlinebb\hfil\line{\hfill\the\footline\hfill}}}
\gdef\doubleformat{\shipout\vbox{\Landspec\makehedlinebb
     \fullline{\box\leftcolumn\hfil\columnbox}\makefotlinebb}
     \advancepageno}
\gdef\columnbox{\leftline{\pagebody}}
\gdef\line#1{\hbox to\hsize{\hskip\leftskip#1\hskip\rightskip}}
\gdef\fullline#1{\hbox to\fullhsize{\hskip\leftskip{#1}%
\hskip\rightskip}}
\gdef\footnote#1{\let\@sf=\empty
         \ifhmode\edef\#sf{\spacefactor=\the\spacefactor}\/\fi
         #1\@sf\vfootnote{#1}}
\gdef\vfootnote#1{\insert\footins\bgroup
         \ifnum\dimnota=1  \eightpoint\fi
         \ifnum\dimnota=2  \ninepoint\fi
         \ifnum\dimnota=0  \tenpoint\fi
         \interlinepenalty=\interfootnotelinepenalty
         \splittopskip=\ht\strutbox
         \splitmaxdepth=\dp\strutbox \floatingpenalty=20000
         \leftskip=\oldssposta \rightskip=\olddsposta
         \spaceskip=0pt \xspaceskip=0pt
         \ifnum\sinnota=0   \textindent{#1}\fi
         \ifnum\sinnota=1   \item{#1}\fi
         \footstrut\futurelet\next\fo@t}
\gdef\fo@t{\ifcat\bgroup\noexpand\next \let\next\f@@t
             \else\let\next\f@t\fi \next}
\gdef\f@@t{\bgroup\aftergroup\@foot\let\next}
\gdef\f@t#1{#1\@foot} \gdef\@foot{\strut\egroup}
\gdef\footstrut{\vbox to\splittopskip{}}
\skip\footins=\bigskipamount
\count\footins=1000  \dimen\footins=8in
\catcode`@=12
\tenpoint
\ifnum\unoduecol=1 \hsize=\tothsize   \fullhsize=\tothsize \fi
\ifnum\unoduecol=2 \hsize=\collhsize  \fullhsize=\tothsize \fi
\global\let\lrcol=L      \ifnum\unoduecol=1
\output{\plainoutput{\ifnum\tipbnota=2 \clearnmbnota\fi}} \fi
\ifnum\unoduecol=2 \output{\if L\lrcol
     \global\setbox\leftcolumn=\columnbox
     \global\setbox\fotlinebb=\line{\hfill\the\footline\hfill}
     \global\setbox\hedlinebb=\line{\hfill\the\headline\hfill}
     \advancepageno  \global\let\lrcol=R
     \else  \doubleformat \global\let\lrcol=L \fi
     \ifnum\outputpenalty>-20000 \else\dosupereject\fi
     \ifnum\tipbnota=2\clearnmbnota\fi }\fi
\def\ifdoublepage{\ifnum\unoduecol=2 }
\gdef\yespagenumbers{\footline={\hss\tenrm\folio\hss}}
\gdef\ciao{ \ifnum\fdefcontre=1 \endfdef\fi
     \par\vfill\supereject \ifnum\unoduecol=2
     \if R\lrcol  \headline={}\nopagenumbers\null\vfill\eject
     \fi\fi \end}
\newskip\olddsposta \newskip\oldssposta
\global\oldssposta=\leftskip \global\olddsposta=\rightskip

 \def\newline{\hfil\break}
\def\jump{\vskip\baselineskip} \newskip\iinnffrr
\def\sjump{\iinnffrr=\baselineskip
          \divide\iinnffrr by 2 \vskip\iinnffrr}
\def\bjump{\vskip\baselineskip \vskip\baselineskip}
\newcount\nmbnota  \def\clearnmbnota{\global\nmbnota=0}
\newcount\tipbnota 

\def\note#1{\global\advance\nmbnota by 1 \ifnum\tipbnota=1
    \footnote{$^{\rm\nttlett}$}{#1} \else {\ifnum\tipbnota=2
    \footnote{$^{\nttsymb}$}{#1}
    \else\footnote{$^{\the\nmbnota}$}{#1}\fi}\fi}
\def\nttlett{\ifcase\nmbnota \or a\or b\or c\or d\or e\or f\or
g\or h\or i\or j\or k\or l\or m\or n\or o\or p\or q\or r\or
s\or t\or u\or v\or w\or y\or x\or z\fi}
\def\nttsymb{\ifcase\nmbnota \or\dag\or\sharp\or\ddag\or\star\or
\natural\or\flat\or\clubsuit\or\diamondsuit\or\heartsuit
\or\spadesuit\fi}   \clearnmbnota
\def\numberfootnote{\global\tipbnota=0} \numberfootnote
\def\setnote#1{\expandafter\xdef\csname#1\endcsname{
\ifnum\tipbnota=1 {\rm\nttlett} \else {\ifnum\tipbnota=2
{\nttsymb} \else \the\nmbnota\fi}\fi} }
\newcount\nbmfig  
\newcount\draftnum \global\draftnum=0
 \def\endformula{\eqno\numero $$}
 \def\efr{\endformula}
\newcount\frmcount \def\clearfrmcount{\global\frmcount=0}
\def\numero{\global\advance\frmcount by 1   \ifnum\indappcount=0
  {\ifnum\cpcount <1 {\hbox{\rm (\the\frmcount )}}  \else
  {\hbox{\rm (\the\cpcount .\the\frmcount )}} \fi}  \else
  {\hbox{\rm (\applett .\the\frmcount )}} \fi}
\def\nameformula#1{\global\advance\frmcount by 1%
\ifnum\draftnum=0  {\ifnum\indappcount=0%
{\ifnum\cpcount<1\xdef\spzzttrra{(\the\frmcount )}%
\else\xdef\spzzttrra{(\the\cpcount .\the\frmcount )}\fi}%
\else\xdef\spzzttrra{(\applett .\the\frmcount )}\fi}%
\else\xdef\spzzttrra{(#1)}\fi%
\expandafter\xdef\csname#1\endcsname{\spzzttrra}
\eqno \hbox{\rm\spzzttrra} $$}
\def\nfr{\nameformula}    \def\numali{\numero}
\def\nameali#1{\global\advance\frmcount by 1%
\ifnum\draftnum=0  {\ifnum\indappcount=0%
{\ifnum\cpcount<1\xdef\spzzttrra{(\the\frmcount )}%
\else\xdef\spzzttrra{(\the\cpcount .\the\frmcount )}\fi}%
\else\xdef\spzzttrra{(\applett .\the\frmcount )}\fi}%
\else\xdef\spzzttrra{(#1)}\fi%
\expandafter\xdef\csname#1\endcsname{\spzzttrra}
  \hbox{\rm\spzzttrra} }      \clearfrmcount
\newcount\cpcount 
\newcount\subcpcount 
\newcount\appcount \def\clearappcount{\global\appcount=0}
\newcount\indappcount \def\clearindappcount{\indappcount=0}
\newcount\sottoparcount 

\def\applett{\ifcase\appcount  \or {A}\or {B}\or {C}\or
{D}\or {E}\or {F}\or {G}\or {H}\or {I}\or {J}\or {K}\or {L}\or
{M}\or {N}\or {O}\or {P}\or {Q}\or {R}\or {S}\or {T}\or {U}\or
{V}\or {W}\or {X}\or {Y}\or {Z}\fi    \ifnum\appcount<0
\immediate\write16 {Panda ERROR - Appendix: counter "appcount"
out of range}\fi  \ifnum\appcount>26  \immediate\write16 {Panda
ERROR - Appendix: counter "appcount" out of range}\fi}
\clearappcount  \clearindappcount \newcount\connttrre
\def\clearconnttrre{\global\connttrre=0} \newcount\countref
\def\clearcountref{\global\countref=0} \clearcountref
\def\references{\goodbreak\null\vbox{\jump\nobreak
   \itemitem{}{\ttaarr References} \nobreak\jump\sjump}\nobreak}
\def\acknowledgements{\goodbreak\null\vbox{\jump\nobreak
\itemitem{ }{\ttaarr Acknowledgements} \nobreak\jump\sjump}\nobreak}
%
%
%
\catcode`@=11
\gdef\Ref#1{\expandafter\ifx\csname @rrxx@#1\endcsname\relax%
{\global\advance\countref by 1    \ifnum\countref>200
\immediate\write16 {Panda ERROR - Ref: maximum number of references
exceeded}  \expandafter\xdef\csname @rrxx@#1\endcsname{0}\else
\expandafter\xdef\csname @rrxx@#1\endcsname{\the\countref}\fi}\fi
\ifnum\draftnum=0 \csname @rrxx@#1\endcsname \else#1\fi}
\gdef\beginref{\ifnum\draftnum=0  \gdef\Rref{\fairef}
\gdef\endref{\scriviref} \else\relax\fi
\ifx\risposta\mplarisposta \ninepoint \fi
\baselineskip=12pt \parskip 2pt plus.2pt }
\def\Reflab#1{[#1]} \gdef\Rref#1#2{\item{\Reflab{#1}}{#2}}
\gdef\endref{\relax}  \newcount\conttemp
\gdef\fairef#1#2{\expandafter\ifx\csname @rrxx@#1\endcsname\relax
{\global\conttemp=0 \immediate\write16 {Panda ERROR - Ref: reference
[#1] undefined}} \else
{\global\conttemp=\csname @rrxx@#1\endcsname } \fi
\global\advance\conttemp by 50  \global\setbox\conttemp=\hbox{#2} }
\gdef\scriviref{\clearconnttrre\conttemp=50
\loop\ifnum\connttrre<\countref \advance\conttemp by 1
\advance\connttrre by 1
\item{\Reflab{\the\connttrre}}{\unhcopy\conttemp} \repeat}
\clearcountref \clearconnttrre
\catcode`@=12
\def\slashchar#1{\setbox0=\hbox{$#1$} \dimen0=\wd0
     \setbox1=\hbox{/} \dimen1=\wd1 \ifdim\dimen0>\dimen1
      \rlap{\hbox to \dimen0{\hfil/\hfil}} #1 \else
      \rlap{\hbox to \dimen1{\hfil$#1$\hfil}} / \fi}
\ifx\oldchi\undefined \let\oldchi=\chi
  \def\cchi{{\raise 1pt\hbox{$\oldchi$}}} \let\chi=\cchi \fi
\def\del{\partial}   
\def\frac#1#2{{\textstyle{#1 \over #2}}}

%
%
%
\newcount\fdefcontre \newcount\fdefcount \newcount\indcount
\global\fdefcontre=0 \global\fdefcount=0 \global\indcount=0
%
%
\null
%
%
%
%
%
%
\loadamsmath
\chapterfont{\bfone} \sectionfont{\scaps}

\def\Teta#1#2{\Theta\left[{}^{#1}_{#2}\right]}

\def\II#1#2{{I}\left[{}^{#1}_{#2}\right]}

\nopagenumbers
{\baselineskip=12pt
\line{\hfill NBI-HE-94-26}
\line{\hfill hep-th/9405100}
\line{\hfill April, 1994}}
{\baselineskip=14pt
\vfill
\centerline{\capsone On the Anomalous Magnetic Moment}
\sjump
\centerline{\capsone in Heterotic Superstrings}
\bjump\bjump
\centerline{\scaps Andrea Pasquinucci and Kaj
Roland~\footnote{$^\dagger$}{Supported by the Carlsberg Foundation.}}
\sjump
\centerline{\sl The Niels Bohr Institute, University of Copenhagen,}
\centerline{\sl Blegdamsvej 17, DK-2100, Copenhagen, Denmark}
\bjump
\vfill
\centerline{\capsone ABSTRACT}
\sjump
\noindent
We explicitly compute the anomalous magnetic moment
at one loop level for an ``electron'' in a 4d
heterotic string theory. The anomalous magnetic moment vanishes if
the model is spacetime supersymmetric, as required by the
supersymmetric sum rules.
\sjump
\vfill
\pageno=0 \eject }
\yespagenumbers\pageno=1
\null\bjump
{\bf 1.} In the string literature there are not too many
explicit computations of loop-amplitudes involving external
spacetime fermions.
Although the tools are known, the technical difficulties involved,
particularly in
four dimensional models, are quite considerable.
On the other hand, amplitudes with external
spacetime fermions are physically very important.
Therefore, it would be of great interest, for example, to generalize
the Bern-Kosower rules [\Ref{BK}] for one-loop amplitudes
with external gauge bosons to include also external spacetime
fermions.

In this letter we consider a much simpler problem: We compute the
1-loop contribution to the anomalous magnetic moment of an
``electron'', i.e. a massive spin 1/2 string state carrying a $U(1)$
charge, using a specific 4-dimensional heterotic string theory.
In a very nice paper, Ferrara and Porrati [\Ref{FP}] obtained,
from the N=1 supersymmetry algebra, the Sum Rules for the
Anomalous Magnetic Moment (AMM) for particles of spin $s$ and mass
$m$.~\note{Similar Sum Rules have been derived for the Electric Dipole
Moment [\Ref{PP}] for models with CP violation.}
For a spin $\frac12$ ``electron" in a $(\frac12, (2) 0)$  multiplet,
they state that the gyromagnetic ratio exactly equals 2, the value one
obtains at the tree level [\Ref{GMR}]. In other
words, the one-loop contribution to the
anomalous magnetic moment should be zero.
This provides us with a non-trivial check on
our computations.

We will thus consider the three-point amplitude of a photon and two
fermions belonging to the first massive level of the string
spectrum, with mass of the order of the Planck mass. We will then
extract from this amplitude the term which contributes to the AMM
(the remaining terms of the amplitude, contributing to the Vertex
Renormalization and to the Electric Dipole Moment,
will be discussed elsewhere). Finally we will
show that this contribution vanishes when the model has
spacetime supersymmetry.

\sjump

{\bf 2.}
We choose to work in the Kawai-Lewellen-Tye (KLT) formalism
for four-dimensional heterotic strings [\Ref{KLT}]. Since this is
based on free world-sheet
fermions it is easy to introduce spin fields describing
the external spacetime fermions. Moreover, we will be able to
consider supersymmetric and non-supersymmetric models at the
same time, since the presence of supersymmetry depends on  the
values of some parameters in the model. Unlike KLT we work in the
Lorentz-covariant formulation, rather than the light-cone gauge.

In a 4-dimensional KLT model we have on top of the four
space-time coordinate fields $X^\mu(z,
\bar{z})$, 22 left-moving complex fermions
$\bar{\psi}_{(\bar{l})}(\bar{z})$, and 10 right-moving
complex fermions $\psi_{(l)}(z)$, the first of which corresponds to the
two transverse space-time Majorana fermions.
Any KLT model is specified by a set of vectors ${\bf W}_i$
defining the set of
possible boundary conditions for the fermions, and a set of
parameters $k_{ij}$
defining the GSO projection. We consider a particular model,
proposed in [\Ref{KLT}], based on the following set of vectors
$$\eqalignno{ & {\bf W}_0\ =\ \left( (\frac12)^{22} \vert
(\frac12) (\frac12\frac12\frac12)^3\right) &\nameali{vectors} \cr
& {\bf W}_1\ =\ \left( (\frac12)^{22} \vert
(0) (0\frac12\frac12)^3\right) \cr
& {\bf W}_2\ =\ \left( (\frac12)^{14} (0)^8 \vert
(0) (0\frac12\frac12) (\frac12 0 \frac12)^2\right) \cr
& {\bf W}_3\ =\ \left( (\frac12)^7 (0)^7 (\frac12)^3 (0)^5 \vert
(0) (\frac12 0 \frac12) (0\frac12\frac12) (\frac12 \frac12 0)\right)
\cr
& {\bf W}_4\ =\ \left( (0)^7 (0)^7 (\frac12)^2 (0) (0)^5\vert
(0) (0\frac12\frac12) (\frac12\frac120) (\frac12\frac12 0)\right)  \ .
\cr}
$$
Each entry in the vector corresponds to a complex fermion, and the spin
structure $\left[ {}^{\alpha_l}_{\beta_l} \right]$
of the fermion $(l)$ is given by
$$\eqalignno{ & \alpha_l  =
\sum_{i=0}^4 m_i ({\bf W}_i)_{(l)} \equiv (m {\bf
W})_{(l)} &  \nameali{spinstructures} \cr
 & \beta_l  =
\sum_{i=0}^4 n_i ({\bf W}_i)_{(l)} \equiv (n {\bf
W})_{(l)} \ ,  \cr }
$$
where the $m_i$ and $n_j$ ($i,j=0,\dots,4$) are
integers with values in $\{0,1\}$. The Ramond (Neveu-Schwarz) sector
corresponds to $\alpha_l = 0\ (\frac12)$.
The constants $k_{ij}$ take
values $0$ or $\frac12$. It turns out that the model is spacetime
supersymmetric if $k_{02} + k_{12} = k_{04} + k_{14}$~mod~1 [\Ref{KLT}].

The model has gauge symmetry $SO(14)\otimes SO(14) \otimes
SO(4) \otimes U(1) \otimes SO(10)$ where the $U(1)$
is associated with the $\overline{17}$ fermion. The once-picture changed
version of the covariant vertex operator for
the $U(1)$ gauge boson is given by
$$
{\cal V}^{(0)} (z,\bar{z};k;\epsilon) = \bar{c}(\bar{z})
\bar\psi^*_{(
\overline{17})}\bar\psi_{(\overline{17})} (\bar{z}) \times c(z)
\left[ \epsilon \cdot \del_z X (z) -i k \cdot
\psi \, \epsilon \cdot \psi (z)\right] e^{ik\cdot X(z,\bar{z})} \ .
\nfr{photon}
Here the superscript $(0)$ refers to the superghost number and
$k_\mu \equiv \sqrt{\alpha^\prime /2}\, p_\mu$ is a dimensionless
momentum satisfying $k^2 = \epsilon \cdot k = 0$. The four Majorana
space-time fermions are related to the two complex ones by
$\psi_\mu= \left\{ \frac1{\sqrt2}(\psi_{(0)} +\psi^*_{(0)}),
\frac1{i\sqrt2}(\psi_{(0)} - \psi^*_{(0)}), \frac1{\sqrt2}(
\psi_{(1)} + \psi^*_{(1)} ), \frac1{i\sqrt2}(\psi_{(1)} -
\psi^*_{(1)})\right\}$, $\mu=0,\dots,3$.
The longitudinal complex fermion, $\psi_{(0)}$, is not
present in the light-cone gauge approach of KLT and does not appear as
an entry in the vectors \vectors. By world-sheet supersymmetry it always
carries the same spin structure as $\psi_{(1)}$.

{}From a detailed study of the spectrum it turns out
that there is a fermion in the ${\bf W}_1 + {\bf W}_3$ sector
with nonzero $U(1)$ charge and mass $1$ (in units of $\alpha^\prime$),
belonging to a $(\frac12, (2) 0)$ multiplet. The vertex operator is
given by
$$\eqalignno{
{\cal V}^{(-1/2)} (z,\bar{z};k;\overline{V},V) \ = \
&\overline{V}^{\bar{A}}
\bar{c}(\bar{z}) \prod_{l=1}^7
S^{(\bar{l})}_{\bar{a}_l} (\bar{z})\prod_{l=15}^{17}
S^{(\bar{l})}_{\bar{a}_l} (\bar{z})\  \times & \nameali{fermion} \cr
& V^A \, c(z) \prod_{l=0,1,4,5,6,7,9} S^{(l)}_{a_l}(z) \,
e^{-\frac12\phi(z)} \ e^{ik\cdot X(z,\bar{z})} \ .  & \cr}
$$
Here the left-moving spinor indices
$\bar{A} = \{\bar{a}_1,\dots,\bar{a}_7;
\bar{a}_{15},\bar{a}_{16};\bar{a}_{17}
\}$ are symmetry indices under the gauge group: The fermion transforms
in the spinor representation of the first $SO(14)$ and the $SO(4)$, and
$\bar{a}_{17} = \pm \frac12$ is the $U(1)$ charge. The right-moving
spinor indices
$A = \{a_0,a_1;a_4,a_5,a_6,a_7,a_9\}$
consist of the 4-dimensional space-time spinor index $\alpha \equiv
(a_0,a_1)$ as well as family indices.
$S^{(l)}_{a_l}$ is the
spin field associated with the $(l)$-th fermion
$$
S_{a_l}^{(l)}(z) = e^{a_l \phi_{(l)}(z)} (c^{(l)})^{a_l} \ ,
\nfr{spinfield}
where $a_l=\pm \frac12$ and
$$
\psi_{(l)}(z) = e^{\phi_{(l)}(z)} c_{(l)} \qquad\qquad\qquad
\psi_{(l)}^*(z) = e^{-\phi_{(l)}(z)} c^*_{(l)}\ .
\nfr{bosonization}
The cocycle factors $c_{(l)}$ guarantee the correct anti-commutation
relations and are given explicitly in ref.~[\Ref{Koste}].
The superghost spin field $e^{-\phi/2}$ also carries a cocycle factor,
which is implicit in eq.~\fermion.

The spinor $\overline{V}^{\bar{A}}$ can be decomposed into a tensor
product
of an $SO(14)$ spinor $\bar{v}_{SO(14)}$, an $SO(4)$ spinor
$\bar{v}_{SO(4)}$ and a 2-component spinor $\bar{v}_{(\overline{17})}$.
Similarly the spinor $V^A$ can be decomposed in
the tensor product of a 4-dimensional spinor $v^\alpha$
and 5 two-dimensional spinors $v_{(l)}^{a_l}$. It satisfies
a Dirac equation which can be obtained
from the requirement $\{ Q_{BRST},{\cal V}^{(-1/2)}\}=0$, given
by
$$\eqalignno{&
V^A(\slashchar{k}_1 + {\Bbb M})_{A}^{\ \, B} =0 &\nameali{dirac} \cr
& {\Bbb M} \equiv \frac12 \Gamma^5 \otimes \sigma_3^{(4)} \otimes \left(
\sigma_1^{(5)} \otimes\sigma_1^{(6)}\otimes\sigma_1^{(7)} +
\sigma_2^{(5)}\otimes\sigma_2^{(6)}\otimes_2^{(7)}\right)\otimes
{\bf 1}^{(9)} \ , \cr}
$$
where $\slashchar{k} \equiv \Gamma^\mu k_\mu \otimes \prod_{l=4,5,6,7,9}
{\bf 1}^{(l)}$ and $\Gamma^\mu$, $\mu=0,\dots,3,5$ are the usual
four-dimensional $\gamma$-matrices whereas the $\sigma^{(l)}_m$,
$m=1,2,3$ are the Pauli matrices acting on the ($l$)-th internal space.

The spinors $\overline{V}^{\bar{A}}$ and $V^A$ are further constrained
by the
set of GSO projections
$$\eqalignno{
& \sigma_3^{(\overline{17})} \sigma_3^{(4)} =
- \exp \left\{ 2\pi i \left[ k_{02} + k_{12} + k_{04} + k_{13} + k_{14}
+ k_{23} + k_{34} \right] \right\} & \nameali{GSO} \cr
&  \sigma_3^{(\overline{1})} \ldots \sigma_3^{(\overline{7})} \Gamma^5
\sigma_3^{(4)} \sigma_3^{(7)} =
\exp \left\{ 2\pi i \left[ k_{00} + k_{01} + k_{02} + k_{03} + k_{12}
+ k_{13} + k_{23} \right] \right\} & \cr
& \sigma_3^{(\overline{15})} \sigma_3^{(\overline{16})} \Gamma^5
\sigma_3^{(4)}
\sigma_3^{(6)} \sigma_3^{(9)} =
\exp \left\{ 2\pi i \left[ k_{00} + k_{01} + k_{03} + k_{04} + k_{13}
+ k_{14} + k_{34} \right] \right\} & \cr
& \Gamma^5 \sigma_3^{(5)} =
- \exp \left\{ 2\pi i \left[ k_{00} + k_{01} + k_{03} + k_{13}
\right] \right\} \ . \cr
}$$
\sjump

{\bf 3.} The starting point for our computation is the on-shell
3-point amplitude on the torus of two
massive ``electrons" \fermion \ and one $U(1)$ gauge boson \photon.
This is given (up to an overall normalization constant) by
$$\eqalignno{  {\cal A}[\epsilon,k ; & V,k_1;U,k_2]\ = \
\sum_{m_i,n_i} C^{m{\bf W}}_{n{\bf W}} \int d^2\tau \int
d^2zd^2z_1 \ \times&\nameali{fundfor} \cr
& \int D[X\psi]D[{\rm (super) ghosts}]\, e^{-S}\,
\vert (\eta_{\tau}\vert b) (\eta_z\vert b) (\eta_{z_1}\vert b)
\vert^2\ \times& \cr
& {\cal V}^{(0)} (z,\bar{z};k;\epsilon) \ \Pi(w) \
{\cal V}^{(-1/2)} (z_1,\bar{z}_1;k_1;\overline{V},V) \ {\cal V}^{(-1/2)}
(z_2,
\bar{z}_2;k_2;\overline{U},U) \ , \cr }
$$
where $\Pi (w)$ is the Picture Changing Operator (PCO)
$$\eqalignno{ &\Pi(w) = 2c\del \xi(w) + 2 e^\phi T^{X,\psi}_F(w) -
\frac12 \del(e^{2\phi}\eta b)(w) -\frac12 e^{2\phi} \del\eta b(w)
&\nameali{pco} \cr
&T^{X,\psi}_F = -\frac{i}2\del X\cdot \psi -\frac{i}2
\sum_{m=1}^2 (\psi^m_{(2)}\psi^m_{(3)}\psi^m_{(4)} + \psi^m_{(5)}
\psi^m_{(6)}\psi^m_{(7)} + \psi^m_{(8)}\psi^m_{(9)}\psi^m_{(10)})\cr
& \psi^m_{(l)} = \left\{ \frac1{\sqrt2}(\psi_{(l)} + \psi^*_{(l)} )\ ,
\ \frac1{i\sqrt2}(\psi_{(l)} - \psi^*_{(l)}) \right\} \ , \cr} $$
and the point $z_2$ has been fixed using the translational invariance of
the
torus. $\eta_{\tau}$, $\eta_z$ and $\eta_{z_1}$ are the Beltrami
differentials dual to the three remaining moduli.
The exact form of the summation coefficients $C^{m{\bf W}}_{n{\bf W}}$
is given in ref.~[\Ref{KLT}].

Following ref.~[\Ref{At}] we have preferred to keep the
PCO at an arbitrary point
$w$, rather than to represent one of the two space-time fermions
by a superghost number $+1/2$ vertex operator (corresponding to, say,
the limit $w \rightarrow z_1$).

As always, the change in the modular integrand induced by a shift in $w$
is a total derivative in the moduli. However, in this simple case the
total derivative vanishes identically due to superghost number
conservation, as may be seen by carrying out the
BRST contour deformation argument of ref.~[\Ref{Verlinde}].
Accordingly, no total derivative ambiguity enters into the modular
integrand.\note{If the second PCO is also inserted at an arbitrary
point, rather than at the position of the photon vertex operator, a
total derivative will appear. But the Lorentz structure of the
ambiguity is such that it will only contribute to the vertex
renormalization, never to the AMM (or the Electric Dipole Moment).}
This is very fortunate, because it means that any vanishing of the
AMM will occur point-by-point in moduli space.

We now turn to the computation of the amplitude. By using the on-shell
conditions, all Lorentz structures appearing in \fundfor\ can be reduced
to three basic ones and the amplitude decomposes into
$$ \eqalignno{
{\cal A}[\epsilon,k ;V,k_1;U,k_2]\ = \ &\epsilon_\mu V{\Bbb M}^2\,
\Gamma^\mu {\Bbb C}U\, {\cal A}_{{\rm VR}}\ + & \nameali{decomp} \cr
&\epsilon_\mu  k_\nu V{\Bbb M}\, \Gamma^{\mu\nu} {\Bbb C}\, U \,
{\cal A}_{{\rm AMM}} +  \epsilon_\mu k_\nu V{\Bbb M}\,
\Gamma^{\mu\nu} \Gamma_5 {\Bbb C}\, U\,
{\cal A}_{{\rm EDM}} \ , \cr} $$
corresponding to Vertex Renormalization, the Anomalous Magnetic
Moment and the Electric Dipole Moment, respectively.
Here ${\Bbb C}$ is the charge conjugation matrix in the
sense that
$$
\Gamma^{\mu} {\Bbb C} = - {\Bbb C} (\Gamma^{\mu})^T \ .
\nfr{cc}
The computation of the correlators appearing in \fundfor\ is in
principle straightforward, but not in practice, and it constitutes the
main effort of this paper.

First, all correlators involving spin fields have been obtained by
bosonization, using the $N$-point 1-loop vertex given in ref.~[\Ref{PDV1}]
for the fermions and in refs.~[\Ref{PDV2}] for the superghosts. The
Lorentz structures in \decomp\ only appear from a careful treatment
of the cocycle factors. Lorentz covariance implies non-trivial
identities among the correlators obtained for different values of the
space-time vector indices. These identities are crucial for simplifying
the final result.
Next, much attention must be given to keeping track of all phases coming
from the GSO projection conditions \GSO.
The final result for the part of the amplitude contributing to the AMM
can be written in the form
$$\eqalignno{
{\cal A}_{\rm AMM} = &{-1\over 2\sqrt2}\left( \bar{v}_{SO(14)}
{\Bbb C}_{SO(14)} \bar{u}_{SO(14)} \right) \left( \bar{v}_{SO(4)}
{\Bbb C}_{SO(4)} \bar{u}_{SO(4)} \right)
\ \times & \nameali{result}\cr
&\left( \bar{v}_{(\overline{17})} \sigma_3^{(\overline{17})}
\sigma_1^{(\overline{17})}
\bar{u}_{(\overline{17})} \right)\,\prod_{i=4,5,6,7,9} \delta_{a_i+b_i,0} \
\sum_{m_i,n_i} K^{m {\bf W}}_{n {\bf W}} \int d^2 \tau  d^2z d^2z_1
\ \times \cr & {1 \over \bar{\omega} (\bar{z}_2) \omega(z_2)} \,
(2\pi {\rm Im} \tau)^{-2}\, \exp \left\{ \frac12
G_B (z_1,\bar{z}_1,z_2,\bar{z}_2) \right\}\,  \bar{{\cal Z}}_L
\, {\cal Z}_R \ {\cal I}_{\rm AMM} \ . &  \cr
} $$
Here ${\Bbb C}_{SO(14)}$ and ${\Bbb C}_{SO(4)}$ are the spinor metrics
of the respective groups, while the factor
$\bar{v}_{(\overline{17})} \sigma_3^{(\overline{17})}
\sigma_1^{(\overline{17})}
\bar{u}_{(\overline{17})}$ is simply the sign of the $U(1)$ charge of
the
``electron''. All phases have been collected in the coefficient
$$\eqalignno{ K^{m{\bf W}}_{n{\bf W}} = &
\exp\left\{2\pi i\left[ \frac12(m_0+n_0) +
\frac12 n_0(m_2+m_3+m_4) \ + \right.\right. &\nameali{sumcoeff} \cr
&\frac12 (n_0 + n_1)(m_2+m_3+m_4) +
\frac12 m_3n_3 + \frac12 m_4n_4 + \frac12 m_3n_4\ +  \cr
& \left( \frac12 + k_{00} + k_{01} + k_{02} + k_{12} + k_{03} + k_{04}
+ k_{23} + k_{24} \right)\, S\ +\cr
& (k_{04}+k_{14}-k_{02}-k_{12})(
S_4 + S_7 + m_4n_0 +m_0n_4+(m_0 + m_1)n_4+\cr
&\left.\left. m_4(n_0 + n_1))\right]\right\} \ , \cr}
$$
where the integer
$$
S_l \equiv (1-2\alpha_l)(1+2\beta_l)
\nfr{sdef}
is even (odd) whenever the $(l)$th spin structure is even (odd), and
$$
S \equiv S_{\overline{17}}+S_4+S_6+S_7 = n_2 m_4 + n_4 m_2 \ ({\rm mod}
\ 2)\ .
\nfr{stot}
$\omega(z)$ is the holomorphic 1-form on the torus, normalized to have
period $2\pi i$ around the $a$-cycle,
and
$$
G_B(z_1,\bar{z}_1,z_2, \bar{z}_2) =
2\left[ \log \vert
E(z_1,z_2)\vert - \frac12 {\rm Re}\left( \int_{z_2}^ {z_1}\omega
\right)^2
{1\over 2\pi {\rm Im}\tau} \right]
\nfr{gb}
is the bosonic Green function. Furthermore,
$$\eqalignno{
\bar{{\cal Z}}_L = &(\bar{\eta} (\bar{\tau}) )^{-24}
\prod_{l=1}^7 \prod_{l=15}^{17} \bar{\Theta} \left[
{}^{\bar{\alpha}_{l}}_{\bar{\beta}_{l}} \right] (\frac12
\bar{\nu}_{12} \vert \bar{\tau} )\ \times & \nameali{zl} \cr
 & \prod_{l=8}^{14} \prod_{l=18}^{22} \bar{\Theta} \left[
{}^{\bar{\alpha}_{l}}_{\bar{\beta}_{l}} \right] ( 0 \vert \bar{\tau} )
\, (\bar{E} (\bar{z}_1,\bar{z}_2))^{-5/2}
\bar{I} \left[
{}^{\bar{\alpha}_{17}}_{\bar{\beta}_{17}} \right]
(\bar{z},\bar{z}_1,\bar{z}_2) \ ,
\cr } $$
$$ \eqalignno{
{\cal Z}_R = &(\eta(\tau))^{-12} {\sqrt{\omega(z_1)
\omega(z_2)}\over \omega (w)} \prod_{l=1,4,5,6,7,9}
\Teta{\alpha_l}{\beta_l} (\frac12 \nu_{12} \vert \tau)\ \times &
\nameali{zr} \cr
& \prod_{l=2,3,8,10} \Teta{\alpha_l}{\beta_l}
(0 \vert \tau)\, (E(z_1,z_2))^{-3/2} \ ,
\cr } $$
where we introduced the Dedekind $\eta$-function
$$
\eta(\tau) = k^{1/24} \prod_{n=1}^{\infty} (1-k^n)\ \, , \qquad
k = e^{2\pi i \tau} \ ,
\nfr{dedekind}
the prime form
$$
E(z_1,z_2) = {2\pi i \Teta00(\nu_{12}
\vert \tau) \over \sqrt{\omega(z_1)\omega(z_2)}
\Teta00^\prime (0\vert \tau) } \ \, , \qquad \nu_{12} =
\int_{z_2}^{z_1} {\omega \over 2\pi i }\ ,
\nfr{primef}
and the shorthand notation
$$\II\alpha\beta(z,z_1,z_2) = \del_z\log{E(z,z_1)\over E(z,z_2)}
+ \left. {2\omega(z)\over 2\pi i} \del_\nu \log \Teta\alpha\beta
(\nu\vert\tau)\right\vert_{\nu= \frac12 \nu_{12}}\ .
\nfr{idef}
Our conventions for the theta functions are
$$\eqalignno{
& \Teta{\alpha}{\beta}(\nu\vert\tau) = e^{i\pi(\frac12-\alpha)^2\tau}
e^{2\pi i(\frac12 +\beta)(\frac12-\alpha)} e^{2\pi i(\frac12
-\alpha)\nu}\ \times & \nameali{theta}
\cr &\qquad \qquad\qquad \prod_{n=1}^\infty
(1-k^n)(1-k^{n+\alpha-1}e^{-2\pi i (\beta+\nu)}) (1-k^{n-\alpha}
e^{2\pi i (\beta+\nu)})\ . \cr} $$
Finally,
$$ \eqalignno{
{\cal I}_{\rm AMM} = &\frac12 (1+ (-1)^S) \times & \nameali{iamm} \cr
& \left\{ \del_zG_B(z,z_1) -
\del_z G_B (z,z_2) - \II{\alpha_1}{\beta_1}(z,z_1,z_2)
\right\} \times &
\cr
& \left\{ \del_wG_B(w,z_1) - \del_w G_B(w,z_2) -
\II{\alpha_L}{\beta_L}(w,z_1,z_2) \right\} \ , \cr } $$
where $L=6$ if $(m_4=m_2, n_4=n_2)$, $L=5$ otherwise.

Only spin structures with $S=0$ (mod 2) contribute to the AMM.
If the definitions \gb \ and \idef \ are inserted into eq.~\iamm \
one finds
that the dependence on $w$ drops out
of the expression \result, as expected.
\sjump

{\bf 4.} Let us consider the case in which the model is supersymmetric.
Looking
at the spectrum, it is easy to recognize how supersymmetry acts. Given
a state in a sector specified by some vector $m{\bf W}$ the
supersymmetric partners are in the sector $m{\bf W} + {\bf W}_0 +
{\bf W}_1$. This follows from the fact that the vector
$$ {\bf W}_0+{\bf W}_1=\left( (0)^{22}\vert (\frac12) (\frac12 00)^3
\right) \quad ({\rm mod\, 1}) \nfr{susy}
is the only one that can be formed from the basis set \vectors\ which
leaves
all left-moving boundary conditions unchanged, that is, can possibly relate
states transforming in the same representation of the gauge group. In fact,
${\bf W}_0+{\bf W}_1$ changes the spin structure of only  four
right-moving
fermions, and clearly interchanges Neveu-Schwarz and Ramond boundary
conditions for the world-sheet supersymmetry current $T^{X,\psi}_F$.

This leads to various conclusions. We want to
show that the AMM vanishes when the model is supersymmetric;
this must be due to the contributions from superpartners circulating
in the loop cancelling one another. Therefore we expect to get zero
by summing, for any given sector $m{\bf W}$, only over the sectors
$m{\bf W}$ and $m{\bf W} + {\bf W}_0 + {\bf W}_1$. We also have to
implement the GSO projections by summing over the $n_i$. However, since
the
$m_i$ and $n_i$ are equivalent by modular invariance, we again expect to
get zero only by summing over boundary conditions $n{\bf W}$ and
$n{\bf W} + {\bf W}_0 + {\bf W}_1$. The explicit calculations vindicate
this belief. This means that it is sufficient to study the right-moving
part of the amplitude. It is
then convenient to change basis for the ${\bf W}$-vectors, or
equivalently,
for the $(m_i,n_i)$. We introduce then $(m^{\prime}_i, n^{\prime}_i)$ by
$$\eqalignno{ & m^{\prime}_1 = m_0 + m_1 \qquad\qquad\qquad m^{\prime}_i
=
m_i \quad i\neq 1 &\nameali{newnm} \cr
& n^{\prime}_1 = n_0 + n_1 \qquad\qquad\qquad\ \  n^{\prime}_i = n_i
\quad
i\neq 1 \ , \cr} $$
since
$$\eqalignno{ m_0{\bf W}_0 + m_1 {\bf W}_1 =\ & m_0 ({\bf W}_0+{\bf
W}_1) + (m_0+m_1) {\bf W}_1  &\numali \cr
=\ & m^{\prime}_0 ({\bf W}_0+{\bf W}_1) + m^{\prime}_1 {\bf W}_1\ ({\rm
mod~1}) \ .\cr} $$
{}From now on we will work in the new basis.

The spin structures depending on $(m_0,n_0)$ are explicitly given by
$$\eqalignno{&
\alpha_1=\frac12 m_0\ ,\quad \alpha_2 = \frac12(m_0+m_3)\ ,\quad
\alpha_5 = \frac12(m_0+m_2+m_4) &\nameali{snomo} \cr
& \alpha_8 = \frac12 (m_0+m_2+m_3+m_4) \cr}
$$
and similarly for the $\beta_i$ in terms of the $n_i$.

In a model with spacetime supersymmetry $k_{02}+k_{12}=k_{04}+k_{14}$
mod~1 and the last term in the summation coefficient \sumcoeff \
drops out so that the dependence on
$(m_0,n_0)$ is given by
$$\eqalignno{
K^{m{\bf W}}_{n{\bf W}}=\exp&\left\{2\pi i\left[ \frac12(m_0+n_0) +
\frac12
n_0(m_2+m_3+m_4)\ \right.\right. & \nameali{summ} \cr
&\quad\left.\left. +\ \hbox{\rm terms not depending on }
(m_0,n_0)\right]
\right\} \ .\cr}
$$
Since we want to sum over $(m_0,n_0)$, keeping all the other integers
fixed, it is convenient to rewrite the theta functions in the (2), (5)
and (8) spaces so that they all have spin structure $(\alpha_1,\beta_1)$.
In doing so, the
argument of each theta function is shifted by
$(\beta- \beta_1) - \tau (\alpha - \alpha_1)$
and we also pick up an overall phase factor
$ \exp \left\{ 2\pi i [ \frac12 n_0
(m_2+m_3+m_4)] \right\} $ cancelling the one appearing in
\summ ,
so that effectively the summation coefficients for the sum over
$(m_0,n_0)$ are reduced to
$$
\exp\left\{2\pi i\left[ \frac12(m_0+n_0) + \ \hbox{\rm
terms not depending on } (m_0,n_0)\right]\right\}\ .
\efr
Thus we expect to prove the vanishing of the AMM in the
supersymmetric case using the standard Riemann identity
$$
\sum_{\alpha,\beta} e^{2\pi i(\alpha+\beta)} \prod_{i=1}^4
\Teta\alpha\beta(x_i\vert\tau) \ =\ 0 \ ,
\nfr{cfor}
where one of the following equations must hold
$$\eqalignno{&x_1+x_2+x_3+x_4 = 0 \qquad\qquad\quad
x_1-x_2-x_3+x_4=0 &\numali\cr
&x_1-x_2+x_3-x_4 = 0 \qquad\qquad\quad x_1+x_2-x_3-x_4 = 0\ .\cr}
$$
Notice that this is not possible in the non-supersymmetric case
since there is an extra dependence on $(m_0,n_0)$ in the phase
of the coefficient \sumcoeff \ given by
$$
(k_{04}+k_{14}-k_{02}-k_{12})(m_4n_0+m_0n_4) = \frac12
(m_4n_0+m_0n_4)\ ({\rm mod}~1)\ .
\efr
\sjump

To show that the AMM vanishes when the model is supersymmetric, it
is convenient to extract all factors depending on $(n_0,m_0)$ from
eqs. \result, \sumcoeff, \zr\ and \iamm, arriving at the quantity
$$\eqalignno{\sum_{m_0,n_0} K^{m{\bf W}}_{n{\bf W}}
&\left( {1\over \log\vert k\vert}
\log \left\vert{z_1\over z_2}
\right\vert + {2\over 2\pi i} \left.\del_\nu \log
\Teta{\alpha_1}{\beta_1}(\nu\vert\tau)\right\vert_{\nu=\frac12 \nu_{12}}
\right)\ \times&\nameali{bfor}\cr
& \left( {1\over \log\vert k\vert} \log\left\vert {z_1\over z_2}
\right\vert + {2\over 2\pi i }\left. \del_\nu \log
\Teta{\alpha_L}{\beta_L}(\nu\vert \tau)\right\vert_{\nu=\frac12
\nu_{12}}
\right)\ \times\cr
& \prod_{l=1,5}\Teta{\alpha_l}{\beta_l}(\frac12 \nu_{12}\vert\tau)
\prod_{l=2,8} \Teta{\alpha_l}{\beta_l} (0\vert\tau)\ .\cr}
$$
Using the Riemann identity eq. \cfor, one can prove that (in the
supersymmetric case)
$$\eqalignno{&\sum_{m_0,n_0} K^{m{\bf W}}_{n{\bf W}} \prod_{l=1,5}
\Teta{\alpha_l}{\beta_l} (\frac12 \nu_{12}\vert\tau) \prod_{l=2,8}
\Teta{\alpha_l}{\beta_l} (0\vert\tau) \ =\ 0 &\numali\cr
&\sum_{m_0,n_0}K^{m{\bf W}}_{n{\bf W}} \left( \left. \del_\nu\log
\Teta{\alpha_1}{\beta_1} (\nu\vert\tau)\right\vert_{\nu=\frac12
\nu_{12}}
+ \left.\del_\nu\log\Teta{\alpha_5}{\beta_5}(\nu\vert\tau)
\right\vert_{\nu=\frac12 \nu_{12}}\right)\ \times\cr
&\qquad\qquad \prod_{l=1,5}\Teta{\alpha_l}{\beta_l}(\frac12
\nu_{12}\vert\tau)
\prod_{l=2,8}\Teta{\alpha_l}{\beta_l} (0\vert\tau)\ =\ 0\cr
&\sum_{m_0,n_0} K^{m{\bf W}}_{n{\bf W}}
\left.\del_\nu\log\Teta{\alpha_1}{\beta_1}
(\nu\vert\tau)\right\vert_{\nu=\frac12 \nu_{12}} \ \left.\del_\nu\log
\Teta{\alpha_5}{\beta_5}(\nu\vert\tau)\right\vert_{\nu=\frac12
\nu_{12}}\
\times\cr
&\qquad\qquad\prod_{l=1,5}\Teta{\alpha_l}{\beta_l}(\frac12
\nu_{12}\vert\tau)
\prod_{l=2,8}\Teta{\alpha_l}{\beta_l}(0\vert\tau)\ =\ 0\ .\cr}
$$
Using these identities it is straightforward to show that the quantity
\bfor\ is zero,
and thus the AMM vanishes due to the spacetime supersymmetry, as it
follows from the sum rules [\Ref{FP}].

In the non-supersymmetric case the quantity \bfor\
is nonzero. Thus to compute the value of the AMM  one
also needs to sum over the other spin structures. This is
particularly difficult since the sum over all the other spin
structures involves more than 4 theta functions and both the
right- and left-movers. Obviously it is always possible to compute
numerically the AMM since the expansion in powers of
$\exp[-2\pi\, {\rm Im} \tau]$ converges very rapidly.
\acknowledgements
We would like to thank Paolo di Vecchia for very interesting
discussions.
\references
\beginref
\Rref{BK}{Z.~Bern and D.~Kosower, Nucl.Phys. {\bf B379} (1992) 451.}
\Rref{FP}{S.~Ferrara and M.~Porrati, Phys.Lett. {\bf B288} (1992) 85.}
\Rref{PP}{M.~Porrati, private communication.}
\Rref{GMR}{S.~Ferrara, M.~Porrati and V.L.~Telegdi, Phys.Rev.{\bf D46}
(1992) 3529.}
\Rref{KLT}{H.~Kawai, D.C.~Lewellen and S.-H.H.~Tye, Nucl.Phys.
{\bf B288} (1987) 1.}
\Rref{Koste}{V.A.~Kostelecky, O.~Lechtenfeld, W.~Lerche, S.~Samuel and
S.~Watamura, Nucl.Phys. {\bf B288} (1987) 173.}
\Rref{At}{J.J.~Atick, L.J.~Dixon and A.~Sen, Nucl.Phys. {\bf B292}
(1987) 109;\newline J.J~Atick and A.~Sen, Nucl.Phys. {\bf B293} (1987)
 317.}
\Rref{Verlinde}{E.~Verlinde and H.~Verlinde, Phys.Lett. {\bf B192}
(1987) 95.}
\Rref{PDV1}{P.~Di Vecchia, M.L.~Frau, K.~Hornfeck, A.~Lerda, F.~Pezzella
and S.~Sciuto, Nucl.Phys. {\bf B322} (1989) 317.}
\Rref{PDV2}{P.~Di Vecchia, invited talk at the ``Workshop on
String Quantum Gravity and Physics at the Planck Energy Scale'', Erice,
June 1992. Published in Erice Theor.Phys. (1992) 16-43;\newline
E.~Verlinde and H.~Verlinde, Nucl.Phys. {\bf B288} (1987) 357.}
\endref
\ciao
